# A systematic investigation of thermal conductivities of transition metal dichalcogenides

Zhongwei Zhang, Yuee Xie, Yulou Ouyang, and Yuanping Chen[*]

*Department of Physics, Xiangtan University, Xiangtan 411105, Hunan, P.R. China*

*Corresponding author*: * chenyp@xtu.edu.cn

**Abstract:** The thermal conductivities of $MoS_2$ and $WS_2$ have been reported by some experimental and theoretical studies, however, the results are different from each other. Here, thermal transport properties of twelve types of single layer transition metal dichalcogenides (TMDs) $MX_2$ (M = Cr, Mo, W; X = O, S, Se, Te) are investigated systematically, by solving Boltzmann transport equation based on first-principle calculations. After accurate considering the size effect and boundary scattering, we find that our calculations can fit the former experimental results well. Moreover, diverse transport properties in TMDs are revealed, and an abnormal dependence of thermal conductivity on atomic mass is observed. In most $MX_2$ structures, the thermal conductivities decrease with the increase of mass of atom M or X. However, the thermal conductivities of sulfides $MS_2$ and selenides $MSe_2$ increase as M changes from Cr to Mo to W, which is contradictory to our traditional understanding. A detailed calculation indicates that the abnormal trend is originated from the rapid increase of phonon relaxation time. Our studies provide important data and effective mechanism for thermal transport in TMDs.

# I. Introduction

The zero band gap of graphene constricts its applications in electronics and optoelectronics, even if it possesses unique physical and chemical properties [1-4]. Therefore, finding semiconductors with suitable band gaps has become a focus in the field of two dimensional (2D) materials [5-9]. The single layer transition metal dichalcogenides (TMDs) are good candidates, which are generally in the form of honeycomb $MX_2$ (M is transition metal while X is chalcogen) [6,7,10]. They exhibit excellent electronic and optoelectronic properties. For example, the carrier mobility of a $MoS_2$ transistor can approach 200 $cm^2$/VS, and its on/off ratio approaches $10^8$ at room temperature [10,11]; high-performance light-emitting transistors and field-effect transistors based on $MoS_2$ and $WS_2$ have also been synthesized [11-14]. Besides the sulfides, transition metal oxides, selenides and tellurides are also found to be semiconductors showing potential applications as electronic and optoelectronic devices [7,15]. Most of the TMDs have been synthesized successfully, which further stimulates extensive interests in the research community of these rich 2D materials [16,17].

Followed by the studies of electronic and optoelectronic properties of TMDs, their thermal transport properties have attracted attentions gradually [18-21], because of crucial roles in the practical applications, such as heat dissipation, phononics and thermoelectric devices. As a typical material of TMDs, the thermal conductivity of $MoS_2$ is numerously investigated by experiment and theory. Zhang *et al.* and Liu *et al.* reported that their measured thermal conductivities are approximate 85 W/mK, with the sample size $d \approx 5$ μm (width) and $L \approx 15$ μm (length) [18,22]. Sahoo *et al.* reported that the thermal conductivity of $MoS_2$ with size $d \approx 4$ μm and $L \approx 10$ μm is 52 W/mK [19]. While Yan *et al.* demonstrated that a smaller circle sample (≈ 1.2 μm) have a much smaller value 34.5 W/mK [23]. Therefore, the experimental values obtained are dependent on the samples. Theoretically calculated thermal conductivities based on various methods present different results, such as 103 W/mK from Boltzmann transport equation [24], 1.5 W/mK [25] and 6.0 W/mK [21] from molecular-dynamics simulations and 23.3 W/mK from the Green's function method [26]. These results not only are different from each other, but also have great disparities with the experimental values. Monolayer $WS_2$ is another typical TMD whose thermal conductivity has been studied [20,24]. A theoretic work revealed that $WS_2$ possesses high thermal conductivity 140 W/mK [24], however, a experimental work reported that the value is only 32 W/mK [20]. Obviously, a systematic study of thermal conductivities of TMDs is necessary, on one hand to clarify the disparities between experimental and theoretical results, on the other hand to obtain a united description of thermal transport properties of the 2D materials.

In this paper, we investigate systematically thermal conductivities of twelve honeycomb $MX_2$ (M = Cr, Mo, W; X = O, S, Se, Te) compounds. In the approaches of dealing with thermal transport, the Green's function method is constricted in the phonon ballistic transport region while molecular-dynamics simulation is strongly related to the classical empirical potentials. Therefore, in many cases, the calculated results based on the two methods deviate from the experimental data. Here, we use Boltzmann transport equation [27-29] combined with first-principle method [30,31] to calculate thermal conductivities of the TMDs. It is found that the thermal transport properties show significant size effect. By considering the boundary scattering and the effect of length, our calculated results agree with the experimental results well. The thermal conductivities of the twelve $MX_2$ compounds as a function of length are predicted. In addition, an abnormal dependence of thermal conductivity on the mass is found: as X = S and Se, the thermal conductivities of MX$_2$ increase with the mass of atom M rather than decrease, which is contradictory to normal understanding. Underlying transport mechanisms are analyzed by heat capacity, phonon group velocity, phonon relaxation time and phonon dispersion.

## II. Computational Method

The thermal conductivity $\kappa$ of TMD is calculated by solving the Boltzmann transport equation, which is expressed as [27-29]:

$$\kappa = \sum C_v v_s(q)^2 \tau_s(q) = \frac{k_b}{4\pi h} \sum_s \int_{q_{min}}^{q_{max}} \frac{\exp[\hbar\omega_s(q)/k_b T]}{\left[\exp[\hbar\omega_s(q)/k_b T]-1\right]^2} q \\ \times [\hbar\omega_s(q)/k_b T]^2 \times v_s(q)^2 \times \tau_s(q) dq \quad , \quad (1)$$

where $s$ and $q$ are phonon branches and wave vectors, respectively; $C_v$, $v_s(q)$ and $\tau_s(q)$ are heat capacity, phonon group velocity and phonon relaxation time, $\omega_s(q)$ is phonon frequency $k_b$ is Boltzmann parameters, $h$ is the thickness of single layer MX$_2$ that chosen as layer distance in bulk structure, and $T$ is the temperature. The $\omega_s(q)$ can be extracted from the phonon spectrum, and the $v_s(q)$ is calculated from $v_s(q) = d\omega_s(q)/dq$. The heat capacity $C_v$ can also be expressed in the frequency presentation, by integrating over the partial phonon density of states ($PDOS(\omega)$) of different particle ($N$) [29,32]:

$$C_v = k_b \sum_N \int_0^\infty \left\{ [\hbar\omega/k_b T]^2 \times \frac{\exp[\hbar\omega/k_b T]}{\left[\exp[\hbar\omega/k_b T]-1\right]^2} PDOS(\omega) \right\} d\omega. \quad (2)$$

To determine the phonon relaxation time $\tau_s(q)$, two types of phonon scattering mechanisms are considered here, including intrinsic phonon-phonon scattering and phonon boundary scattering. The intrinsic phonon-phonon scattering process is studied by considering the three-phonon Umklapp scattering [27,33]:

$$\tau_s^U(q) = \frac{1}{\gamma_s(q)^2 q^2} \frac{M v_s(q)^2}{k_B T} \frac{\omega_{s,\max}}{\omega_s()^2}, \tag{3}$$

where $\omega_{s,\max}$ is the maximum frequency of branch $s$, $m$ is the total mass of atoms in a unit cell, $\gamma_s(q)$ is Grüneisen parameters that can be obtained from the phonon spectrum by $\gamma_s(q) = -V d\omega_s(q)/\omega_s(q)dV$ [26,34]. The phonon boundary scattering can be evaluated as [27,29]:

$$\tau_s^B(q) = \frac{d}{v_s(q)} \frac{1+p}{1-p}, \tag{4}$$

where $d$ is the width of a sample, and $p$ is the specularity parameter which characterizes the fraction of specularity of scattered phonons depending on the roughness of the edge, ranging from 0.0 for a completely rough edge to 1.0 for a perfectly smooth edge. Then, according to the Matthiessen's rule, $\tau_s(q)$ can be found by combine the phonon-phonon Umklapp scattering and phonon boundary scattering together [35]:

$$\frac{1}{\tau_s(q)} = \frac{1}{\tau_s^U(q)} + \frac{1}{\tau_s^B(s,q)}. \tag{5}$$

It is noted that the phonon mean free path cannot exceed the physical length $L$ of the sample, and thus the phonons should be excluded if they cannot satisfy the following formula [28],

$$v_s(q)\tau_s^U(q) < L. \tag{6}$$

The first principle calculations for phonon spectra are accurately performed by the Vienna ab initio simulation package (*VASP*) based on density functional theory [30]. The generalized gradient approximation with Perdew-Burke-Ernzerhof exchange-correlation potential (GGA-PBE) is adopted, with plane wave cut-off energy 450 eV. The total energy convergence is chosen as $1.0 \times 10^{-8}$ eV, and structure optimization is performed until the force acting on each atom is less than $1.0 \times 10^{-7}$ eV. In addition, the calculations are performed on a $5 \times 5 \times 1$ supercell structure.

## III. Results and Discussion

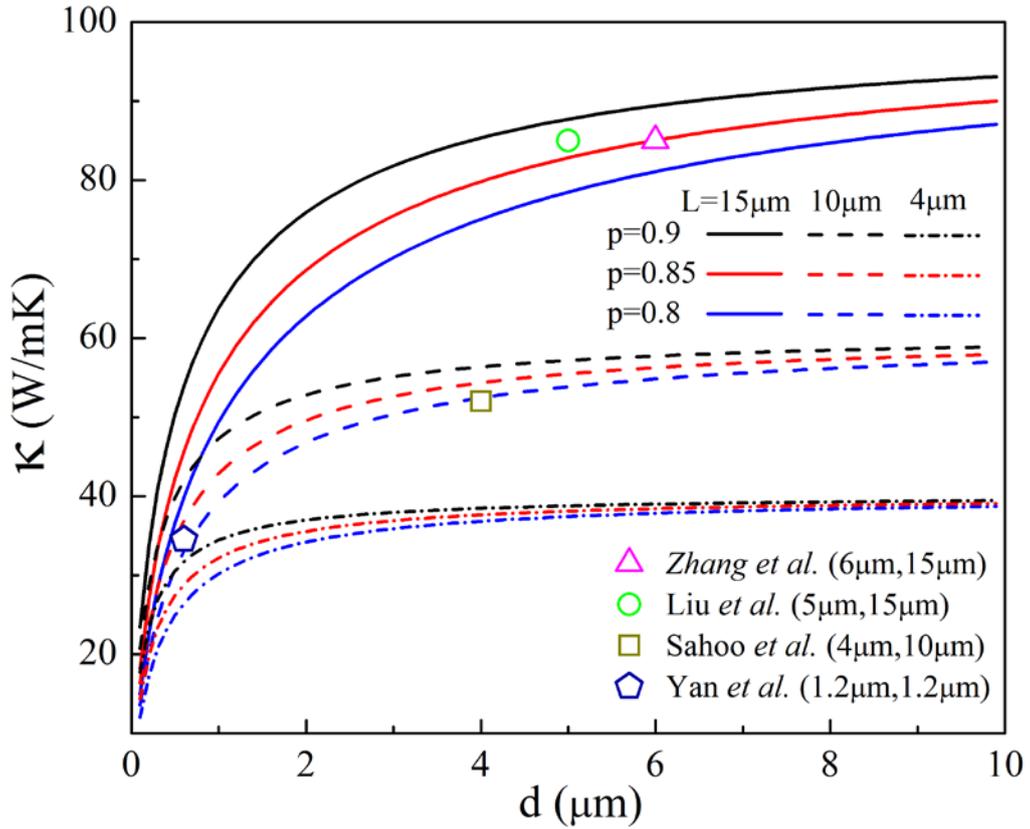

**Figure 1** Thermal conductivity $\kappa$ of MoS$_2$ as a function of width $d$ with different $p$ and length $L$, respectively, at $T = 300$K. The single symbols are experimental results from *Zhang et al.* in Ref

[18], *Liu et al.* in Ref [22], *Sahoo et al.* in Ref [19] and *Yan et al.* in Ref [23] with different sample sizes ($d$, $L$).

In Fig. 1, the effects of width $d$ and length $L$ on the thermal conductivity $\kappa$ of MoS$_2$ are shown. To describe the boundary scattering in detail, the effect of $p$ is also considered. Three length values $L$ = 4, 10 and 15 μm, corresponding to sample lengths in four experiments, are selected. If $L$ is fixed, $\kappa$ increases rapidly at small $d$ followed by convergence at large $d$ ($d \gtrsim L$). The effect of $p$ also becomes weak with the increase of $d$. This is reasonable because the influences of the boundary scattering are weakened in the samples with larger width. Meanwhile, one can find that $\kappa$ exhibits significant length dependence: the longer $L$, the higher $\kappa$. We compare our calculations with the former reported experimental results shown as single symbols. The calculated results are agreement well with the experimental reports. Therefore, the differences between the experimental values are mainly because of the distinct sample sizes. When $p$ is in the range of 0.8 ~ 0.9, the calculated results can fit the experimental results better, indicating that there exists weak specularity of scattered phonons because of rough edges. To accurately simulate the phonon transport in the experimental samples, in the following calculations, $p$ is set as 0.85.

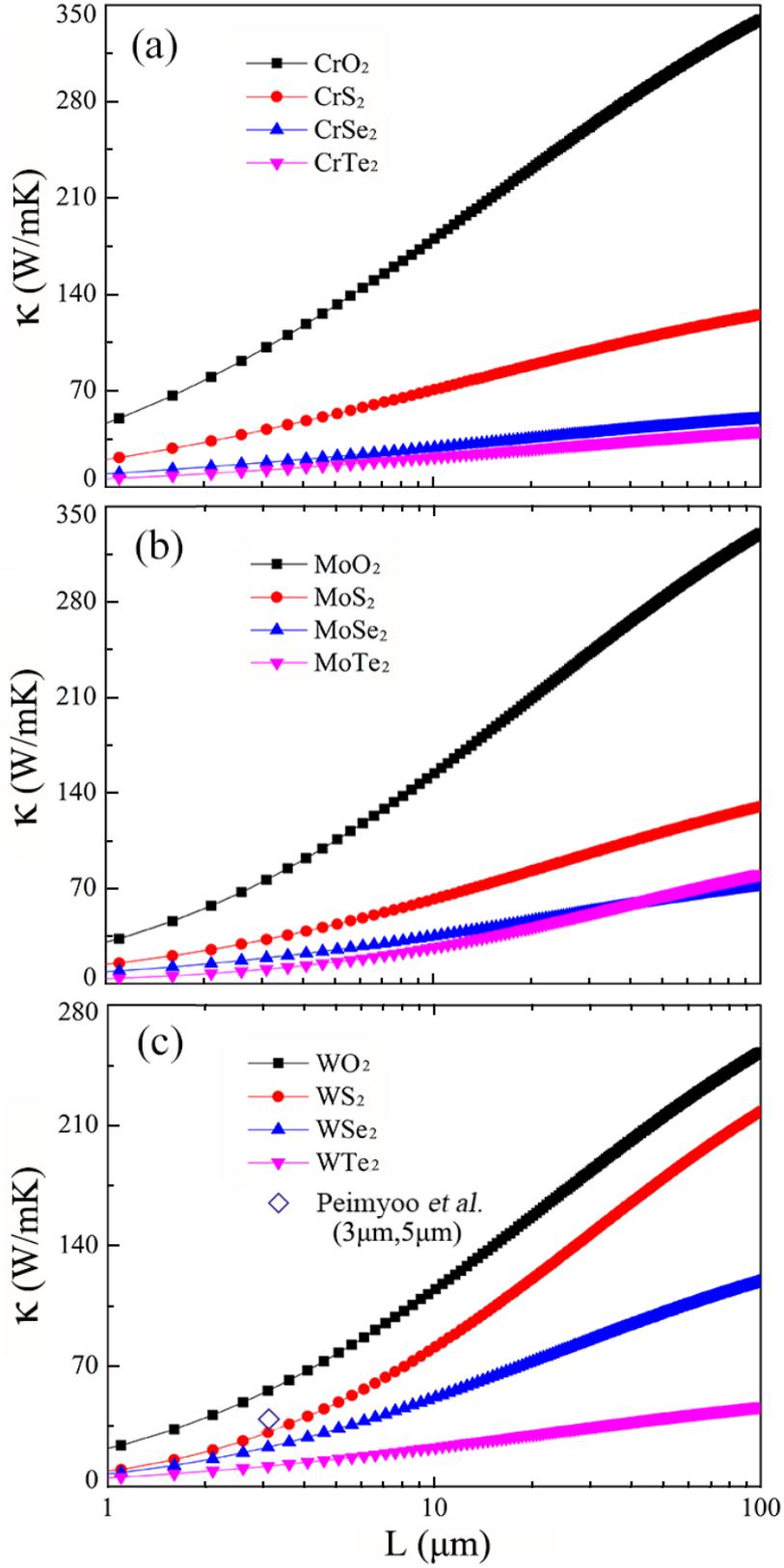

**Figure 2** Thermal conductivities $\kappa$ of (a) CrX$_2$, (b) MoX$_2$, and (c) WX$_2$ (X = O, S, Se and Te) as functions of $L$, respectively. Other parameters are fixed as $d$ = 5 μm, $p$ = 0.85 and T = 300 K. The single symbol is the reported thermal conductivity of WS$_2$ with sample size ($d$, $L$) from Peimyoo *et al.* in Ref [20].

The $\kappa$ values of twelve MX$_2$ (M = Cr, Mo, W; X = O, S, Se, Te) as functions of $L$ at the room temperature are calculated systematically ($d$ = 5 μm), as shown in Fig. 2. The curves illustrate that the $\kappa$ of these materials have different length dependent behaviors. Relatively, the $\kappa$ values of transition metal oxides and WS$_2$ increase faster with $L$ than those of other TMDs. As $L$ changes from 1 to 100 μm, the $\kappa$ values of oxides and WS$_2$ ascend approximately 10 times, while others only 4 ~ 5 times. It is noted that all the $\kappa$ values approach to convergence at $L$ = 100 μm, i.e., equal to the values of TMDs with infinite $L$. Peimyoo *et. al* reported that $\kappa$ of a WS$_2$ sample with $L$ = 3 μm and $d$ = 5 μm is 31.5 W/mK [20]. Our calculation is agreement well with the experimental value, as shown in Fig. 2(c). Seen from Fig. 2, the $\kappa$ values of all TMDs are lower, especially transition metal selenides and tellurides. However, TMDs have been proved that possess high electron mobility [36,37], and thus should be excellent thermoelectric materials [38,39]. In Table I, the $\kappa$ values of the twelve MX$_2$ with $d$ = 5 μm and $L$ = 15 μm at the room temperature are presented. These results would benefit for understanding thermal transport in thermoelectric devices and heat dissipation ability in nanoelectronic devices based on TMDs.

| MO$_2$ | $\kappa$(W/mK) | MS$_2$ | $\kappa$(W/mK) | MSe$_2$ | $\kappa$(W/mK) | MTe$_2$ | $\kappa$(W/mK) |
|---|---|---|---|---|---|---|---|
| CrO$_2$ | 236.2 | CrS$_2$ | 88.8 | CrSe$_2$ | 35.3 | CrTe$_2$ | 27.9 |
| MoO$_2$ | 209.6 | MoS$_2$ | 82.2 | MoSe$_2$ | 46.2 | MoTe$_2$ | 41.4 |
| WO$_2$ | 157.5 | WS$_2$ | 121.2 | WSe$_2$ | 72.7 | WTe$_2$ | 30.2 |

**TABLE 1** Thermal conductivity $\kappa$ of twelve MX$_2$ (M = Cr, Mo, W; X = O, S, Se, Te) compounds in the size of $d$ = 5μm and $L$ = 15μm. Other parameters are fixed at $p$ = 0.85 and $T$ = 300K.

In Table 1, one can find that the variation of $\kappa$ exhibits two trends. One is $\kappa$ drops gradually with the increase of atom mass, for example, the chalcogens in CrX$_2$,

MoX$_2$ and WX$_2$ vary from O to Te. This is a normal phenomenon because heavy atoms weaken the vibrations and thus drop $\kappa$ value. However, there exists another abnormal trend in Table 1. For example, in MS$_2$, MSe$_2$ and MTe$_2$, $\kappa$ do not drop monotonously when M vary from Cr to Mo to W. Especially, the $\kappa$ of MSe$_2$ increase with the atomic mass, which is contrary to the traditional understandings. In fact, the abnormal trend not only occurs at some specific lengths. Figure 3(a) shows a three dimensional plot for the $\kappa$ values of TMDs with a large (or infinite) $L$. Solid blue lines guide our eyes to view the variation of transition metals while dashed red lines for the variation of chalcogens. Abnormal trend is observed in MS$_2$ and MSe$_2$ whose $\kappa$ increase with the masses of transition metal atoms.

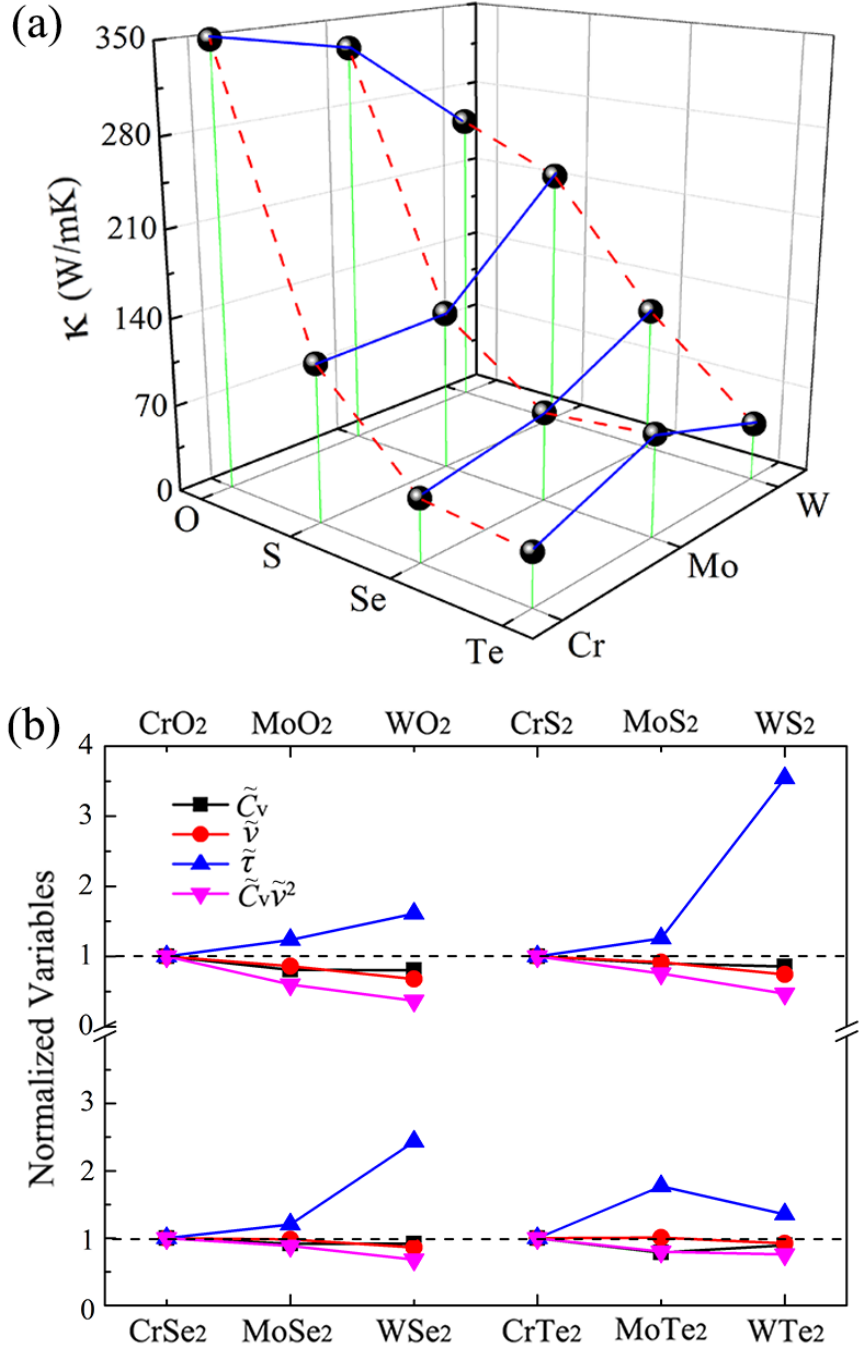

**Figure 3** (a) Schematic view of thermal conductivities $\kappa$ of twelve $MX_2$ (M = Cr, Mo, W; X = O, S, Se, Te) compounds, at $T = 300$ K. Other parameters are fixed as $d = 5$ μm and $p = 0.85$. (b) The normalized variables $\tilde{C}_v$, $\tilde{v}$, $\tilde{\tau}$ and $\tilde{C}_v\tilde{v}^2$ of $MX_2$ with the variation of transition metal M.

According to Eq. (1), $\kappa$ is determined by three factors, $C_v$, $v_s(q)$ and $\tau_s(q)$. To explore the underlying reasons of the abnormal trend, in Fig. 3 (b), we calculate the average values of the factors $\tilde{C}_v$, $\tilde{v}$ and $\tilde{\tau}$ for MX$_2$ with the variation of M. A normalized scales are used here to show a better comparison. One can find that, both $\tilde{C}_v$ and $\tilde{v}$ drop as the transition metal M change from Cr to Mo to W, and $\tilde{C}_v \tilde{v}^2$ further reveal the descending trends. While most of $\tilde{\tau}$ rise with the masses of transition metals except transition metal tellurides. Therefore, the variation of $\kappa$ is determined by the competition of $\tilde{C}_v \tilde{v}^2$ and $\tilde{\tau}$. Relatively, in sulphides and selenides, the variation of $\tilde{\tau}$ with the masses of M is more drastic than that of $\tilde{C}_v \tilde{v}^2$, which results in the abnormal trend of $\kappa$. We also compare $\tilde{C}_v$, $\tilde{v}$ and $\tilde{\tau}$ of CrX$_2$, MoX$_2$ and WX$_2$ (X = O, S, Se and Te). The variations of $\tilde{C}_v$ and $\tilde{v}$ dominate the variation of $\kappa$, and thus a normal trend of $\kappa$ is observed in these compounds. The details are shown in Fig. S1 in Supporting Material (SM).

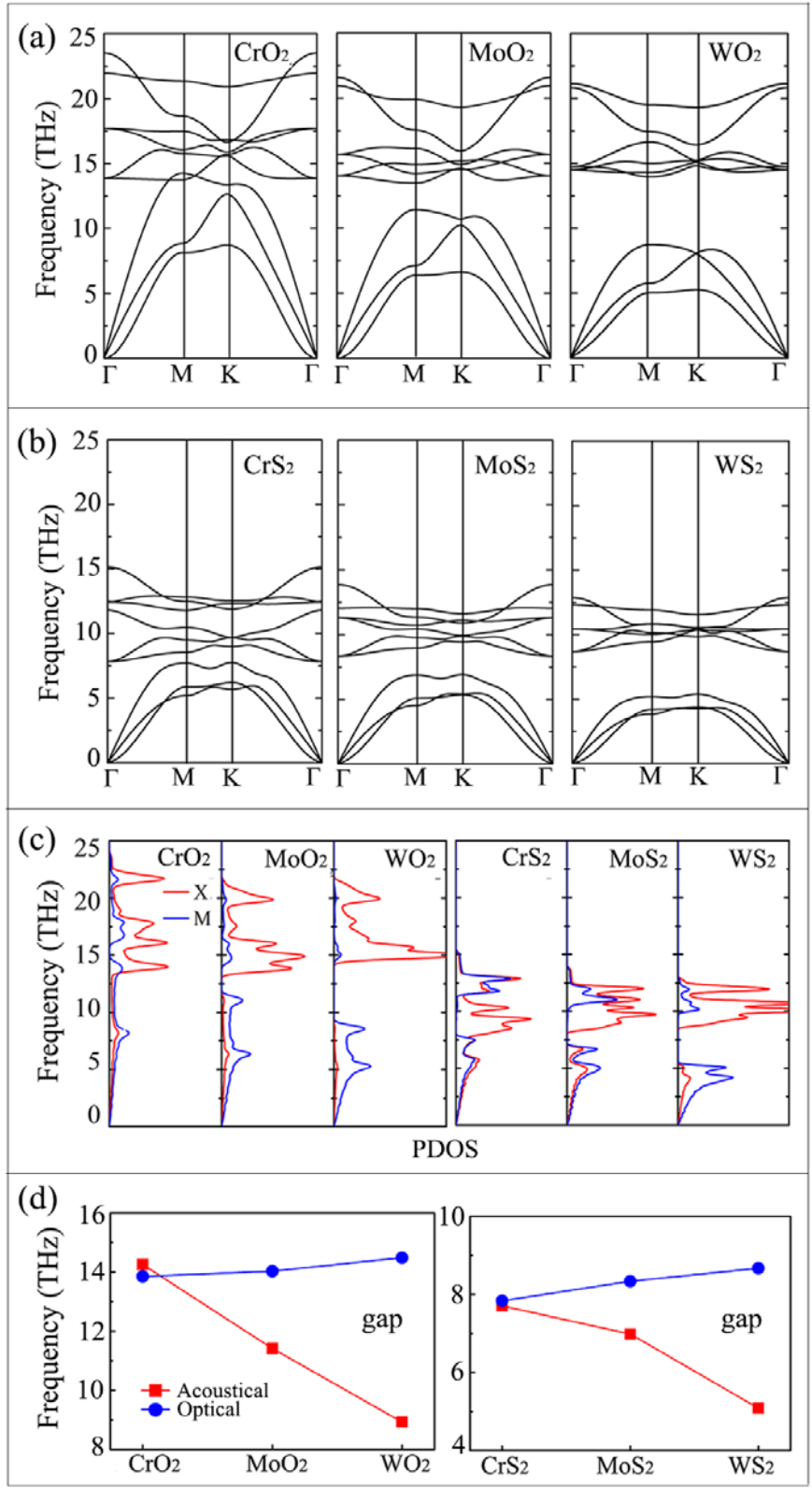

**Figure 4** Phonon dispersions of (a) $MO_2$ and (b) $MS_2$ (M = Cr, Mo and W). (c) PDOS and (d) band gap between acoustical and optical branches of $MO_2$ and $MS_2$ with the variation of transition metal M.

To further explore the $\kappa$'s abnormal trend in oxides and sulphides, in Fig. 4, we compare their phonon dispersions, PDOS and band gaps. Figures 4(a) and 4(b) present the phonon dispersions of $MO_2$ and $MS_2$ (M = Cr, Mo and W), respectively. One can find that the frequency ranges of acoustical phonons in $MO_2$ are larger than those of $MS_2$, and the variation ranges of the former are also larger than those of the latter as M varies from Cr to W. Therefore, $\tilde{v}$ of oxides decreases faster than that of sulphides. The comparison of PDOS profiles in Fig. 4(c) shows that the variation of PDOS is similar to the case of $\tilde{v}$, and thus the $\tilde{C}_v$ of oxides decreases faster than that of sulphides as M varies from Cr to W, according to Eq. (2). In Fig. 4(d), we present the band gaps between acoustical and optical branches of oxides and sulfides. The gap prohibits some important phonon scattering channels, and thus $\tilde{\tau}$ is enlarged significantly [24,40]. One can find that the variation ratios of band gap for oxides and sulfides are different, which results in different increase velocity of $\tilde{\tau}$. Therefore, the abnormal trend of sulfides is because on one hand the slower decrease of $\tilde{C}_v$ and $\tilde{v}$ on the other hand the faster increase of $\tilde{\tau}$.

## IV. Conclusion

In summary, we have carried out a systematic investigation of thermal conductivity of twelve $MX_2$ compounds (M = Cr, Mo, W; X = O, S, Se, Te) by using Boltzmann transport equation combined with first-principle method. By considering the effects of boundary scattering and geometric sizes, our calculation results fit the former experimental values well. We also find that oxides and $WS_2$ possess relatively high heat dissipation abilities, while other compounds show low thermal conductivities. As the transition metal M varies from Cr to Mo to W, the thermal conductivities of $MS_2$ and $MSe_2$ exhibit an abnormal mass dependence, which is explained by the variation of $\tilde{C}_v$, $\tilde{v}$ and $\tilde{\tau}$.

## Acknowledgments

This work was supported by the National Natural Science Foundation of China (Nos. 51376005 and 11474243).